\documentclass[aps,preprint]{revtex4-1} 
\usepackage{graphicx}
\usepackage{amssymb}
\usepackage{epstopdf}
\usepackage{amsmath}
\usepackage{bm}

\newcommand{\half}{\frac{1}{2}}

\newcommand{\dd}{\partial}

\newcommand\refeq[1]{(\ref{#1})}
\newcommand\reffig[1]{Figure~\ref{#1}}
\newcommand\of[1]{\left( #1 \right)}

\newcommand\vecbf[1]{{\bf #1}}

\newcommand\hatx{\hat{\vecbf x}}
\newcommand\mat[1]{{\mathbb #1}}

\begin{document}

\title{Classical and Quantum Mechanical Motion in Magnetic Fields}

\author{J. Franklin}
\email{jfrankli@reed.edu}
\author{K. Cole Newton}
\affiliation{Department of Physics, Reed College, Portland, Oregon 97202,
USA}

\begin{abstract}
We study the motion of a particle in a particular magnetic field configuration both classically and quantum mechanically.  For flux-free radially symmetric magnetic fields defined on circular regions, we establish that particle escape speeds depend, classically, on a gauge-fixed magnetic vector potential, and demonstrate some trajectories associated with this special type of magnetic field.  Then we show that some of the geometric features of the classical trajectory (perpendicular exit from the field region, trapped and escape behavior) are reproduced quantum mechanically using a numerical method that extends the norm-preserving Crank-Nicolson method to problems involving magnetic fields.   While there are similarities between the classical trajectory and the position expectation value of the quantum mechanical solution, there are also differences, and we demonstrate some of these.

\end{abstract}

\maketitle

\section{Introduction}

There is a well-known problem (Problem 5.43) in~\cite{Griffiths} that asks the reader to show that if a charged particle starts at the center of a circular (of radius $R$), radially-symmetric, flux-free magnetic field region, it will exit the region (if it exits) perpendicular to the circular boundary.  This is an exercise in angular momentum conservation, and its ultimate utility resides in running the problem backwards:  if you shoot a particle into a region with this special magnetic field, it will hit the center provided it enters perpendicular to the circular boundary of the region.

Our interest in the problem begins with the determination of the critical velocity that allows the particle to escape the field region at all.  Since there is no traditional potential energy barrier to go over, it is not immediately obvious what sets the minimum ``escape" speed here.  After we determine the condition for escape, highlighting the role of a gauge-fixed magnetic vector potential in classical mechanics, we turn to particle trajectories in quantum mechanics.

From Schr\"odinger's equation in a region with magnetic vector potential $\vecbf A$, we can establish that the expectation value of position satisfies the following ODE (see~\cite{GriffithsQM} Problem 4.59, for example~\cite{NOTEONE}):
\begin{equation}\label{expp}
m \, \frac{d^2 \langle \vecbf x \rangle}{d t^2} = \frac{q}{2 \, m} \, \langle \vecbf p \times \vecbf B - \vecbf B \times \vecbf p \rangle - \frac{q^2}{m} \, \langle \vecbf A \times \vecbf B \rangle,
\end{equation}
where $\vecbf p = m \, \vecbf v + q \, \vecbf A \dot = \frac{\hbar}{i} \, \nabla$ is the canonical momentum.  If the magnetic field was constant, this would reduce to
\begin{equation}\label{wrongexpp}
m \, \frac{d \langle \vecbf v \rangle}{d t} = q \, \langle \vecbf v\rangle \times \vecbf B,
\end{equation}
and the expectation value $\langle \vecbf v \rangle$ would be directly comparable to the classical velocity.  For magnetic fields that are {\it not} constant, the right-hand side of~\refeq{expp} defines an exotic effective force, one which is very different from $q \, \langle \vecbf v \rangle \times \vecbf B$ (as we shall see).  As an equation of motion, we don't know what to expect for $\langle \vecbf x \rangle$ from~\refeq{expp}.  Indeed, we shall see that the expectation value of position is quite different from the classical position vector for these magnetic trajectories, and there are other differences as well.  If the equation of motion for the expectation value of position was~\refeq{wrongexpp}, we would expect the ``speed" (the magnitude of $\langle \vecbf v \rangle$ here) to be constant, just as it is classically.  But the effective force on the right of~\refeq{expp} does not lead to a constant magnitude for the expectation value of velocity.

There are also similarities between the classical trajectories and the position expectation value of quantum mechanical solutions -- we will use a numerical solution of Schr\"odinger's equation to show that the expectation value of kinetic energy is constant (as it should be for motion in a magnetic field), and we can also establish that certain geometric properties of the quantum mechanical trajectory are shared with the classical trajectory:  the particle exits the field region perpendicular to the boundary, and we can get both ``bound" motion, and ``escape" trajectories.  The difference between the trajectory-based ``speed", $\sqrt{\langle \vecbf v \rangle \cdot \langle \vecbf v \rangle}$ and the kinetic energy ``speed", $\sqrt{\langle \vecbf v \cdot \vecbf v \rangle}$ is the main distinction between the classical and quantum mechanical trajectories, but it is a significant difference.

\section{Escape Speed}

The Lagrangian for a particle moving in the presence of a magnetic field is:
\begin{equation}
L = \half \, m \, \vecbf v \cdot \vecbf v + q \, \vecbf v \cdot \vecbf A,
\end{equation}
where $\vecbf A$ is the magnetic vector potential.  The canonical momentum is then $\vecbf p \equiv \frac{\dd L}{\dd \vecbf v} = m \, \vecbf v + q \, \vecbf A$.  The Legendre transform of the Lagrangian defines the Hamiltonian:
\begin{equation}\label{HofA}
H = \vecbf v \cdot \vecbf p - L = \frac{1}{2 \, m} \, \of{\vecbf p - q \, \vecbf A} \cdot \of{\vecbf p - q \, \vecbf A}.
\end{equation}
We know that the Hamiltonian is conserved, and that the speed of the particle is also a constant of the motion (typical of motion in magnetic fields, which do no work).

For our target problem, the magnetic field points in the $\hat{\vecbf z}$ direction, and we're interested in motion occurring in the $x-y$ plane (we will set the initial velocity to lie in this plane).  The magnetic vector potential takes the general form: $\vecbf A = A(s) \, \hat{\bm \phi}$ (its magnitude depends only on $s$, similar to the magnetic field itself).  In polar coordinates, the Hamiltonian is
\begin{equation}
H = \frac{1}{2 \, m} \, \left[ p_s^2 + \frac{1}{s^2} \, p_\phi^2 + q^2 \, A^2 - 2 \, \frac{q}{s} \, p_\phi \, A\right],
\end{equation}
and we can immediately identify the conserved $p_\phi$ (angular momentum) from the equation of motion: $\dot p_\phi = -\frac{\dd H}{\dd \phi} = 0$.  The magnetic vector potential acts as a momentum, and we have to be careful to separate the velocity portion of the canonical momentum, $m \, \vecbf v$ (with its constant magnitude), from the potential part.  In order to untangle the two, at least initially, we'll take $A(0) = 0$, and give the particle initial speed $v_0$ (in the $\hatx$ direction).  Since we are starting at the origin, we'll pick the constant $p_\phi = 0$ (to avoid the $1/0^2$ and $1/0$ that would appear in $H$ otherwise).

Under these simplifying (but reasonable) assumptions, the initial value of the Hamiltonian is:
\begin{equation}
E = \frac{1}{2} \, m \, v_0^2,
\end{equation}
just the kinetic energy of the particle at $t = 0$.  At any other time,  we have
\begin{equation}
E = \frac{1}{2 \, m} \, \left[ p_s^2 + q^2 \, A^2 \right],
\end{equation}
so that
\begin{equation}\label{psvA}
p_s = \pm \sqrt{ (m \, v_0)^2 - (q \, A)^2}.
\end{equation}
Because of the form of $\vecbf A$ (which points in the $\hat{\bm \phi}$ direction), the radial momentum is $p_s = m \, \dot s$, and we can solve~\refeq{psvA} for $\dot s$,
\begin{equation}\label{dots}
\dot s = \pm \sqrt{v_0^2 - \of{\frac{q \, A}{m}}^2}.
\end{equation}
The value of $\dot s$ cannot be imaginary (when $\dot s = 0$, all of the motion occurs in the $\hat{\bm \phi}$ direction), and so this relation provides precisely the desired ``escape speed" -- if a particle is to exit the field region, it must have
\begin{equation}\label{escape}
v_0 \ge \frac{q \, A_{\hbox{\tiny{max}}}}{m}
\end{equation}
where $ A_{\hbox{\tiny{max}}}$ is the maximum vector potential magnitude over the domain.

What do we make of the fact that if we take $v_0$ less than this escape speed, there will be imaginary values for $\dot s$?  The particle never gets to those regions -- when $\dot s = 0$, the particle turns around, so that all of the motion will occur within a circle of radius $\bar s$ defined by the value of $A$ at which $v_0 = \frac{q \, A(\bar s)}{m}$.  The escape speed in~\refeq{escape} uses the maximum value of $A$ in order to overcome all such constraining circles.

\section{Flux-free fields}

The escape speed depends on the magnitude of the vector potential, but the vector potential has gauge freedom, how do we know that the maximum ``height" is being pinned down to a unique value?  So far, we have required that $\vecbf A = A(s) \, \hat{\bm \phi}$, appropriate for a radially symmetric magnetic field pointing in the $\hat{\bf z}$ direction, in Coulomb gauge. We also took $A(0) = 0$ in order to set the initial particle angular momentum to zero.

For flux-free fields (over the domain of the disk of radius $R$), there is an additional requirement:
\begin{equation}
0 = \int \vecbf B \cdot d\vecbf a = \oint \vecbf A \cdot d{\bm \ell},
\end{equation}
and for our form for $\vecbf A$, this reads
\begin{equation}
0 = \int_0^{2\pi}\,  A(R) \, R \, d\phi = 2 \, \pi \, R \, A(R),
\end{equation}
which means that $A(R) = 0$.  What could we add to $\vecbf A$ that preserves these basic requirements?  The gradient of a function $f$ could be added to $\vecbf A$, $\vecbf A \rightarrow \vecbf A + \nabla f$, yielding the same magnetic field.  But, if we are to remain in Coulomb gauge, $f$ must be a harmonic function, $\nabla^2 f = 0$.  The flux-free boundary condition imposes the additional requirement that $f$ is independent of $\phi$ (else we can't get $\nabla f = 0$ at $R$ for all $\phi$), so we are left with $f = a \, \log(s/s_0)$ for constant $a$ and $s_0$, which will not allow us to set the boundary condition at $s = 0$ ($\nabla f \vert_{s =0} = 0$) unless $a = 0$.   So in this case, the gauge is fully fixed, and that's what allows us to unambiguously identify an escape speed.

We can also use this $A(R) = 0$ requirement to solve the original problem posed in~\cite{Griffiths} -- from~\refeq{dots}, we have, at $R$: $\dot s = v_0$, so that all of the velocity is in the $\hat{\vecbf s}$ direction, with none of it in the $\hat{\bm \phi}$ direction, the particle exits the region radially.

\section{Example}

As a model flux-free, radial magnetic field, confined to the region $s \le R$, take
\begin{equation}
\vecbf B = \left\{ \begin{array}{ll} B_0 \, \of{1 - \frac{3 \, s}{2 \, R} } \, \hat{\vecbf z} & s \le R \\
0 & s > R \end{array} \right.,
\end{equation}
this linear magnetic field is the simplest we can pick that can be made flux-free.  
The potential that satisfies the requirements $A(0) = A(R) = 0$, and whose curl matches $\vecbf B$ is
\begin{equation}\label{Avec}
\vecbf A = \left\{ \begin{array}{ll} \frac{B_0 \, s}{2} \, \of{1 - \frac{s}{R}} \, \hat{\bm \phi} & s \le R \\
0 & s > R \end{array} \right., 
\end{equation}
predictably quadratic in $s$.  The first term in the parentheses represents a constant magnetic field of magnitude $B_0$.

\begin{figure}[htbp] 
   \centering
   \includegraphics[width=2.5in]{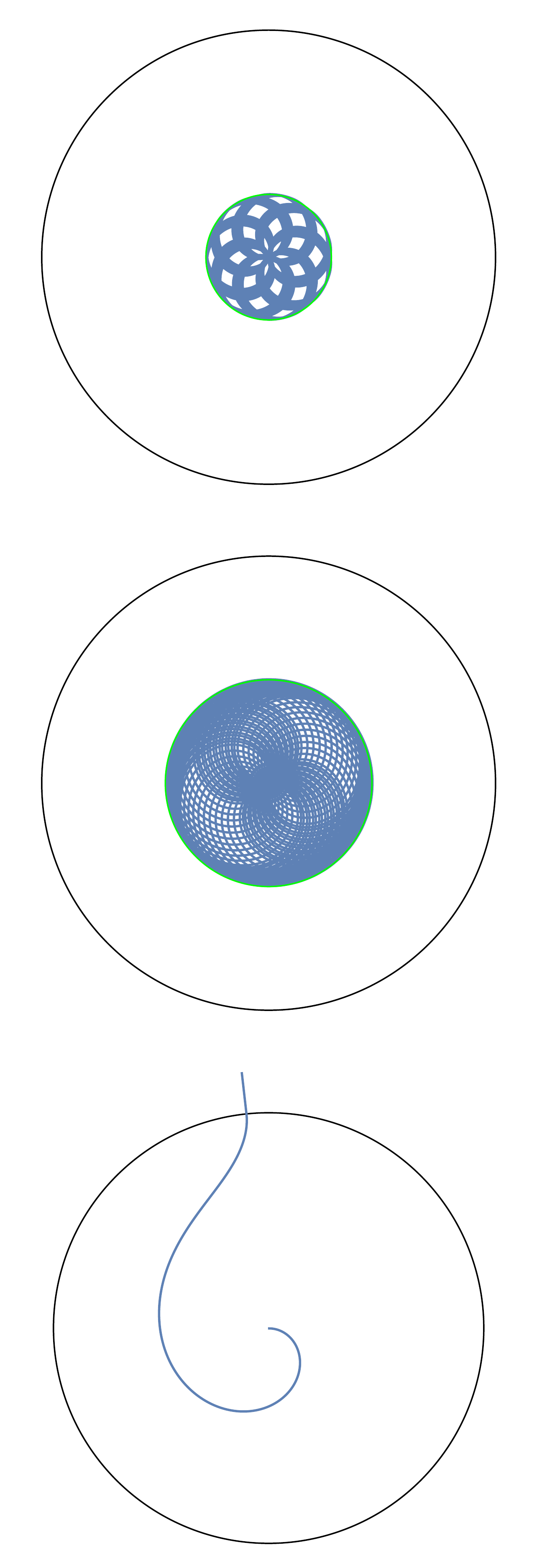} 
   \caption{Trajectories of particles moving in a linear, flux-free field -- the initial speed is increasing from top to bottom.  In the top two, the radius of the circle that bounds the trajectory has been calculated (by solving $q \, A(\bar s)/m = v_0$ for $\bar s$) and is shown in green.  For the bottom plot, $v_0$ is above the escape speed, and the particle exits perpendicular to the boundary.}
   \label{fig:exe}
\end{figure}

Here, we can determine the escape speed, $q \, A_{\hbox{\tiny max}}/m$, analytically -- the maximum of the potential occurs at $s = R/2$ where the magnitude is $B_0 \, R/8$.  For initial speeds less than this, we will get bound trajectories, and for initial speeds greater than this, the particle will exit the field region perpendicular to the boundary.  Examples are shown in~\reffig{fig:exe}, in which we plot two bound trajectories together with their bounding circles (of radius $\bar s$ obtained by solving $v_0 = \frac{q \, A(\bar s)}{m}$ for $\bar s$), and the trajectory for a particle that escapes.  These trajectories were generated using a standard fourth-order Runge-Kutta solver.

\section{Quantum Mechanics}

On the quantum mechanical side, we start with the same Hamiltonian~\refeq{HofA} in Schr\"odinger's equation
\begin{equation}
\frac{1}{2 \, m} \, \of{\vecbf p - q \, \vecbf A} \cdot \of{\vecbf p - q \, \vecbf A} \, \Psi = i \, \hbar \, \frac{\dd \Psi}{\dd t},
\end{equation}
where we understand that $\vecbf p = \frac{\hbar}{i} \, \nabla$.  Writing out Schr\"odinger's equation with the momentum substitution in place,
\begin{equation}
\frac{1}{2 \, m} \, \left[ -\hbar^2 \, \nabla^2 \, \Psi + i \, \hbar \, q \, \nabla \cdot (\vecbf A \, \Psi) + i \, \hbar \, q \, \vecbf A \cdot \nabla\, \Psi+ q^2 \, A^2\, \Psi \right]  = i \, \hbar \, \frac{\dd \Psi}{\dd t},
\end{equation}
let $\nabla = \ell_0^{-1}\, \bar{\nabla}$, $\vecbf A = A_0 \, \bar{\vecbf A}$, $t = t_0 \, \bar t$, where the barred variables are dimensionless, then
\begin{equation}\label{dimproblem}
\frac{1}{2} \, \left[ -\bar{\nabla}^2 \, \Psi + i \, \alpha \, \of{\bar{\nabla} \cdot (\bar{\vecbf A} \, \Psi) + \bar{\vecbf A} \cdot \bar{\nabla} \, \Psi} + \alpha^2 \, \bar A^2\, \Psi \right] = i \, \frac{\dd \Psi}{\dd \bar t}
\end{equation}
for $\frac{\hbar \, t_0}{m \, \ell_0^2} = 1$, and where $\alpha \equiv \frac{q \, A_0 \, t_0}{m \, \ell_0} = \frac{q \, A_0 \, \ell_0}{\hbar}$ is a dimensionless variable that allows us to set the magnitude of the vector potential.

Our starting point will be a Gaussian centered at the origin with initial momentum expectation value $\langle \vecbf p \rangle = p \, \hatx$ -- normalized and written in Cartesian coordinates:
\begin{equation}\label{initialGaussian}
\Psi_0(x,y) = a \, \sqrt{\frac{2}{\pi}} \, e^{-a^2 \, (x^2 + y^2)} \, e^{i \, p \, x/\hbar},
\end{equation}
where $a$ is a parameter that tells us how sharply peaked the Gaussian is -- the standard deviation for this initial Gaussian is $\sigma = \frac{1}{2 \, a}$.
Using $x = \ell_0 \, \bar x$, $y = \ell_0 \, \bar y$, $p = m \, \ell_0/t_0 \, \bar p$, $a = \bar a/\ell_0$, the initial wavefunction can be written in terms of the dimensionless variables,
\begin{equation}
\Psi_0 = \frac{1}{\ell_0} \,  \bar a \, \sqrt{\frac{2}{\pi}} \, e^{-\bar a^2 \, (\bar x^2 + \bar y^2)} \, e^{i \, \bar p \, \bar x}
\end{equation}
with $\frac{m \, \ell_0^2}{\hbar \, t_0} = 1$ from above.  We can, finally, introduce the dimensionless wave function: $\Psi = \frac{1}{\ell_0} \, \bar\Psi$, where the initial $\bar \Psi_0$ is just the above with the factor of $\ell_0^{-1}$ removed.

\subsection{Numerical Method}

We'll use a norm-preserving modification of Crank-Nicolson, developed in~\cite{Newton}.  The idea is to use finite difference to generate forward and backward Euler methods (as with the usual Crank-Nicolson, see, for example~\cite{CN}) but in a way that preserves the Hermiticity of the discrete Hamiltonian.  To define the elements of the method, introduce a grid in (the dimensionless) $\bar x$ and $\bar y$:
$\bar x_j = j \, \Delta$ and $\bar y_k = k \, \Delta$ for constant spacing $\Delta$.  We'll also discretize in time, $\bar t_n = n \, \Delta \bar t$.  Let $\bar \Psi^n_{jk} = \bar \Psi(\bar x_j, \bar y_k, \bar t_n)$, with $\bar A^x_{jk} = \bar A^x(\bar x_j, \bar y_k)$ and similarly for $\bar A^y_{jk}$ (the magnetic vector potential is time-independent here).  Using finite difference approximations to the derivatives in~\refeq{dimproblem}, with a forward Euler approximation for the temporal derivative gives
\begin{equation}\label{method}
\begin{aligned}
\bar \Psi^{n+1}_{jk} &= \bar \Psi^n_{jk} + 
\frac{i \, \Delta \bar t}{2}\,  \biggl[\frac{\bar\Psi^n_{(j+1)k} - 2 \, \bar \Psi^n_{jk} + \bar\Psi^n_{(j-1)k}}{\Delta^2} + \frac{\bar \Psi^n_{j(k+1)} - 2 \, \bar \Psi^n_{jk} + \bar \Psi^n_{j(k-1)}}{\Delta^2} \\
-&i \, \alpha \, \biggl( \of{ \frac{\bar A^x_{(j+1) k} \, \bar \Psi^n_{(j+1) k} - \bar A^x_{(j-1) k} \, \bar \Psi^n_{(j-1) k}}{2 \, \Delta} + \frac{\bar A^y_{j (k+1)} \, \bar \Psi^n_{j(k+1)} - \bar A^y_{j(k-1)} \, \bar\Psi^n_{j(k-1)}}{2 \, \Delta} } \\
& + \bar A^x_{jk} \, \of{\frac{\bar \Psi^n_{(j+1)k} - \bar \Psi^n_{(j-1) k}}{2\, \Delta} } +  \bar A^y_{jk} \, \of{\frac{\bar \Psi^n_{j(k+1)} - \bar \Psi^n_{j(k-1)}}{2\, \Delta} }   \biggr) 
- \alpha^2 \, \of{ \of{\bar A^x_{jk}}^2 + \of{\bar A^y_{jk}}^2} \,\bar \Psi^n_{jk} 
\biggr].
\end{aligned}
\end{equation}

Suppose our spatial grid has $N$ points in both the $\bar x$ and $\bar y$ directions, then we can embed the spatial values of $\bar \Psi$, at time level $n$, in a vector of length $N^2$:
\begin{equation}
{\bm \Psi}^n \dot\equiv \left( \begin{array}{c} \bar \Psi^n_{11} \\ \bar\Psi^n_{21} \\ \vdots \\ \bar \Psi^n_{N1} \\ \bar \Psi^n_{12} \\ \bar \Psi^n_{22} \\ \vdots \end{array} \right),
\end{equation}
so that given the $\bar x$ and $\bar y$ grid locations, $j$ and $k$ (respectively), the index in ${\bm \Psi}^n$ is: $g(j,k) = (k-1) \, N + j$.
Using these spatial vectors, the Euler update above can be written as a matrix-vector product, defining $\mat H$ from the details of the right-hand-side of~\refeq{method}:
\begin{equation}
{\bm \Psi}^{n+1} = \of{ \mat{I} + i \, \Delta \bar t \, \mat H} \, {\bm \Psi}^n.
\end{equation}
From its definition, $\mat H^\dag = \mat H$, it is Hermitian by construction.
Similarly, backwards Euler takes the form:
\begin{equation}
\of{ \mat I - i \, \Delta \bar t \, \mat H} \, {\bm \Psi}^{n+1} = {\bm \Psi}^n,
\end{equation}
and the Crank-Nicolson method is then defined by
\begin{equation}
\of{ \mat I - i \, \frac{1}{2} \, \Delta \bar t \, \mat H} \, {\bm \Psi}^{n+1} =\of{ \mat{I} + \frac{1}{2} \, i \, \Delta \bar t \, \mat H} \, {\bm \Psi}^n.
\end{equation}
This method is norm-preserving, and can be used with our initial wave function, projected onto the grid, to develop the $n^{th}$ update:
\begin{equation}
{\bm \Psi}^n =\of{ \left[  \mat I - i \, \frac{1}{2} \, \Delta \bar t \, \mat H \right]^{-1} \, \left[ \mat{I} + \frac{1}{2} \, i \, \Delta \bar t \, \mat H\right] }^n \, {\bm \Psi}^0.
\end{equation}
Implicit in the method is that the wave function must be zero at the boundary of the numerical domain (that allows us to set the values of $\bar\Psi$ at the boundaries, the $0$ and $N+1$ points, in~\refeq{method}) -- our problem is immersed in an infinite square box in Cartesian coordinates.

\section{Comparison}

We chose to make the spatial grid with $N = 200$ points in each direction, extending from $-10$ to $10$ (in dimensionless length).  Our (dimensionless) time step was $\Delta\bar t = .01$, and we took $\bar p = 4$ in the initial wave function -- that tells us roughly how many steps it would take to get the position expectation value of a free Gaussian to hit the edge of the domain: $\sim 250$ steps.  In order to probe the behavior inside the field region, we took $\bar R = 2$ so that a free Gaussian's position expectation value would leave the region in $\sim 50$ time steps. To choose $\bar a$, note that the standard deviation for a free Gaussian is
\begin{equation}
\sigma = \sqrt{\frac{1}{4 \, \bar a^2} + \bar a^2 \, \bar t^2},
\end{equation}
and we would like the rate of spreading to be small compared to the expectation value of momentum, so that roughly: $\bar a < \bar p$.  The initial expectation value of momentum is numerically determined (even though $\bar p$ is specified, we may or may not capture it numerically), and that determination is sensitive to the choice of $\bar a$ -- if the initial Gaussian is too sharply peaked, there will not be enough representation on the grid to numerically integrate the expectation value accurately.  We found that $\bar a = 1$ led to $\langle \bar p \rangle = 3.86$, an initial error of $\sim 4 \%$ (given that $\bar p =4$) due to: 1.\ the finite difference approximation to the derivative (needed to approximate $\frac{\hbar}{i} \, \nabla$), and 2.\ the use of a simple box-sum to approximate the expectation value integrals.  The choice $\bar a = 1$ also localized the particle inside the field region -- the probability of finding the particle within the circle of radius $\bar R$ was, numerically, $.9997$ at $\bar t = 0$.

\begin{figure}[htbp] 
   \centering
   \includegraphics[width=3in]{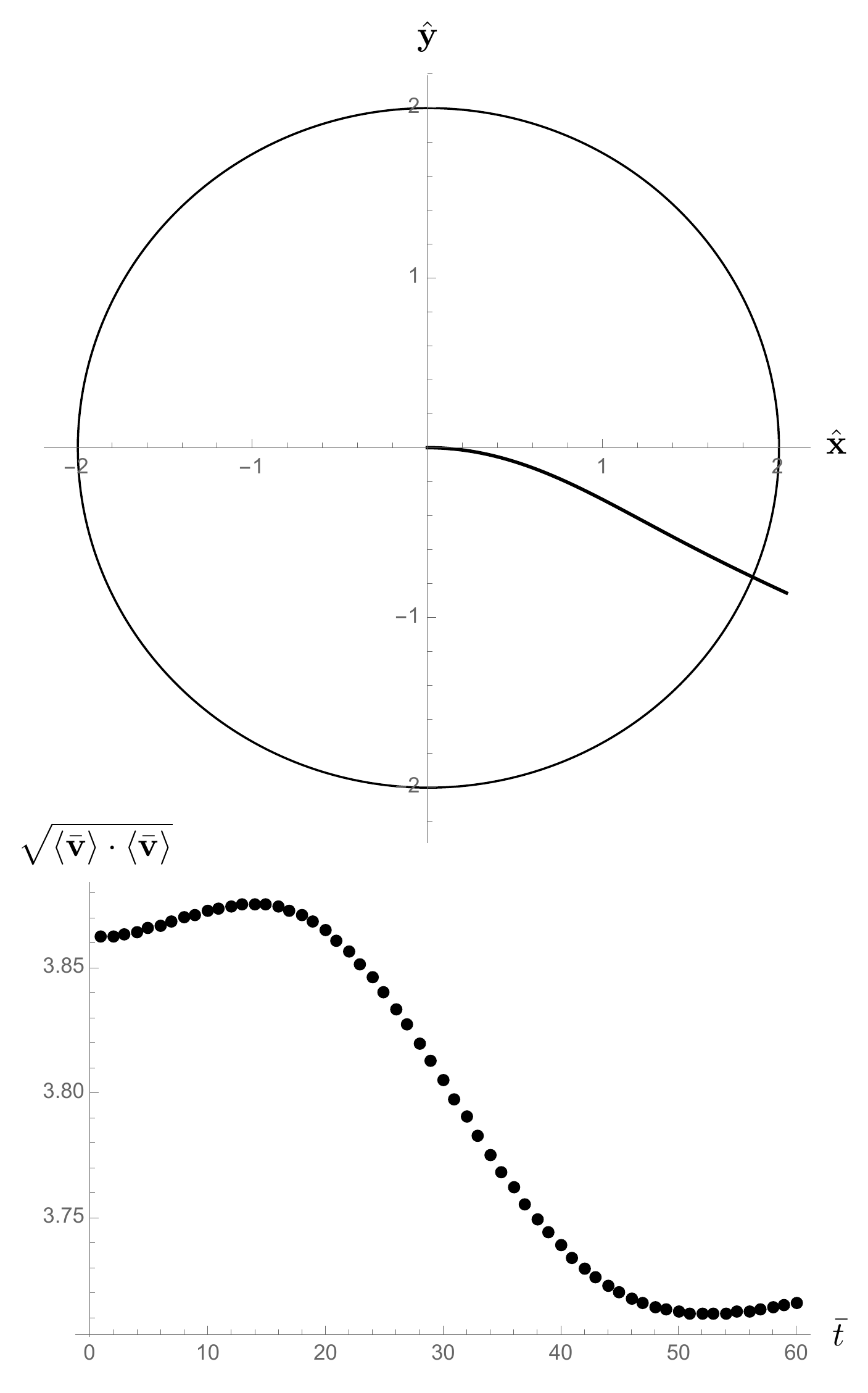} 
   \caption{The trajectory (expectation value) of a particle (top) -- the field region is within the circle of radius $\bar R = 2$.  The ``speed" of the particle as a function of time is shown below.  }
   \label{fig:exit}
\end{figure}

With these choices in place, we used the Crank-Nicolson method described above to move the initial Gaussian forward in time with $\alpha = 5$.  The method preserved norm very well -- the difference between the max and min total probability over the time of numerical solution was $\sim 10^{-13}$.   After running for $\sim 60$ steps, the expectation value of position indicated that the particle had left the field region, and a plot of that exit is shown in~\reffig{fig:exit}.  The velocity vector at exit makes an angle of $\sim 1.6$ (radians) with $\hat{\bm \phi}$ at the location of exit, so the velocity vector is roughly perpendicular to the boundary, with an error of $\sim 2 \%$ (equivalent in size to the initial error in the expectation value of momentum).  The expectation value of energy $\langle E \rangle = \int \Psi^* \, H \, \Psi \, d\tau$  (calculated numerically using finite differences for derivatives and a simple box sum for the integration) has max-minus-min value of $10^{-14}$ over the first $60$ times steps, so that energy is conserved well here.

We also calculate the expectation value of the particle's velocity: $\langle \bar {\vecbf v} \rangle \equiv \frac{d \langle \bar{\vecbf x}\rangle}{d \bar t}$ (using finite difference to approximate the time-derivative), and from that we can compute the ``speed" of the particle (the magnitude of $\langle \vecbf v \rangle$) -- that is also shown in~\reffig{fig:exit}.  The speed is not constant, but difference over the range in question is still within $\sim 4\%$, so it is not clear if this is just the original error or if the speed is truly fluctuating.  

\begin{figure}[htbp] 
   \centering
   \includegraphics[width=4in]{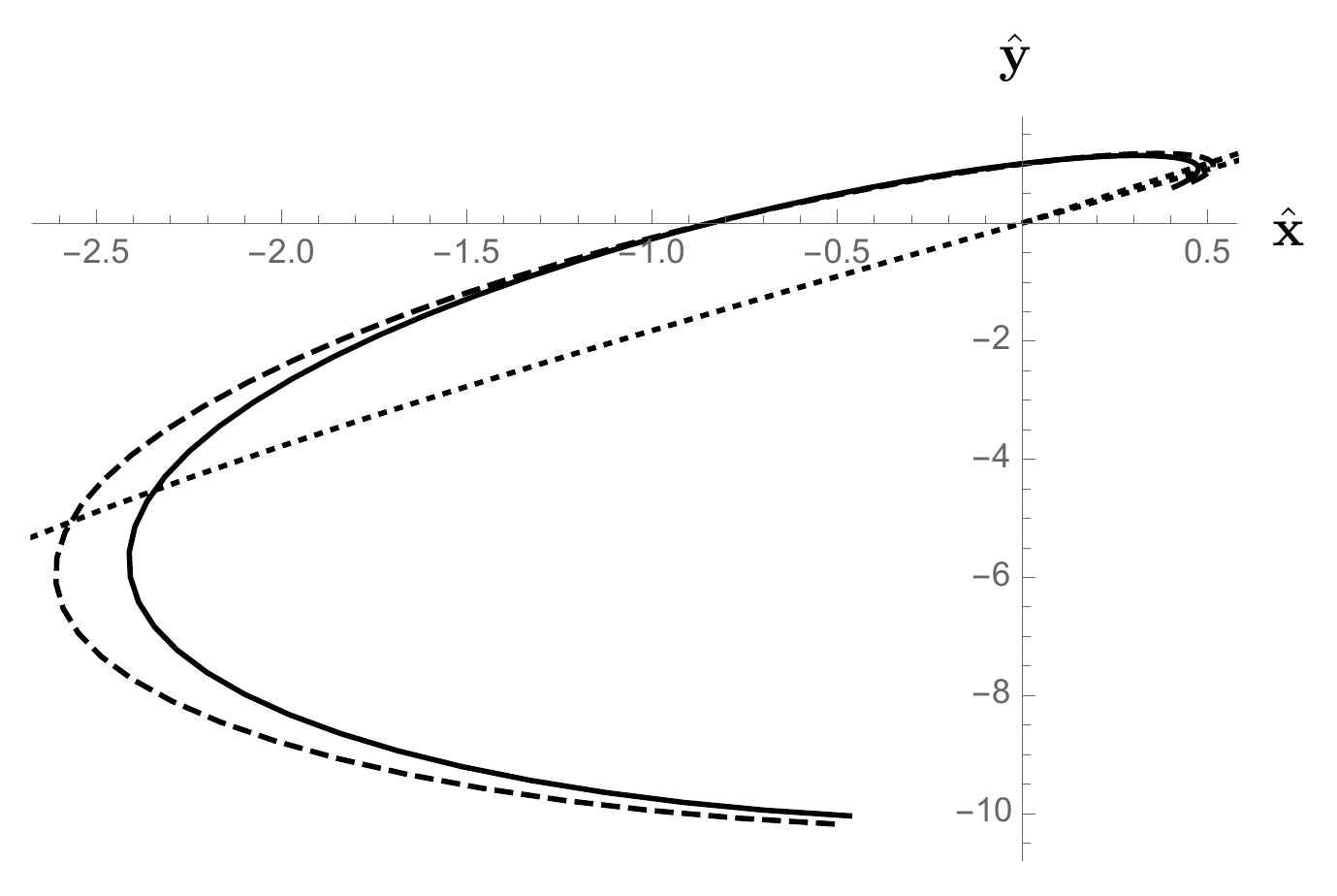} 
   \caption{The ``force" (the curves here connect the tips of the force vectors) associated with the quantum mechanical trajectory -- the solid line is calculated from the approximate second time-derivative of the position expectation value $\langle \vecbf x \rangle$ (and represents the left-hand side of~\refeq{expp}), the dashed line is the expectation value found on the right-hand side of~\refeq{expp}, and the dotted line is the value of $q \, \langle \vecbf v \rangle \times \vecbf B$.}
   \label{fig:allots}
\end{figure}

From the expectation value of position, we can also generate $\frac{d^2 \langle \vecbf x \rangle}{d t^2}$ using finite difference for the temporal derivative, and we can compare that with the effective force defined by the right-hand-side of~\refeq{expp}.  We can also establish that the effective force defined by the right-hand-side of~\refeq{wrongexpp} (namely $q \, \langle \vecbf v \rangle(t) \times \vecbf B(\langle \vecbf x \rangle(t))$) is not the one generating the motion here by computing it explicitly -- in~\reffig{fig:allots}, we plot the left-hand-side of~\refeq{expp} as a function of time (the curve shown in the plot connects the tips of these force vectors), together with the effective forces from the right-hand-sides of~\refeq{expp} and~\refeq{wrongexpp}.
It is clear that while the correspondence between the left and right-hand sides of~\refeq{expp} (the solid and dashed lines in~\reffig{fig:allots}) is not perfect, the effective force defined by~\refeq{expp} is far closer to governing the dynamics of $\langle \vecbf x \rangle$ than the effective force defined by~\refeq{wrongexpp}.

\begin{figure}[htbp] 
   \centering
   \includegraphics[width=3in]{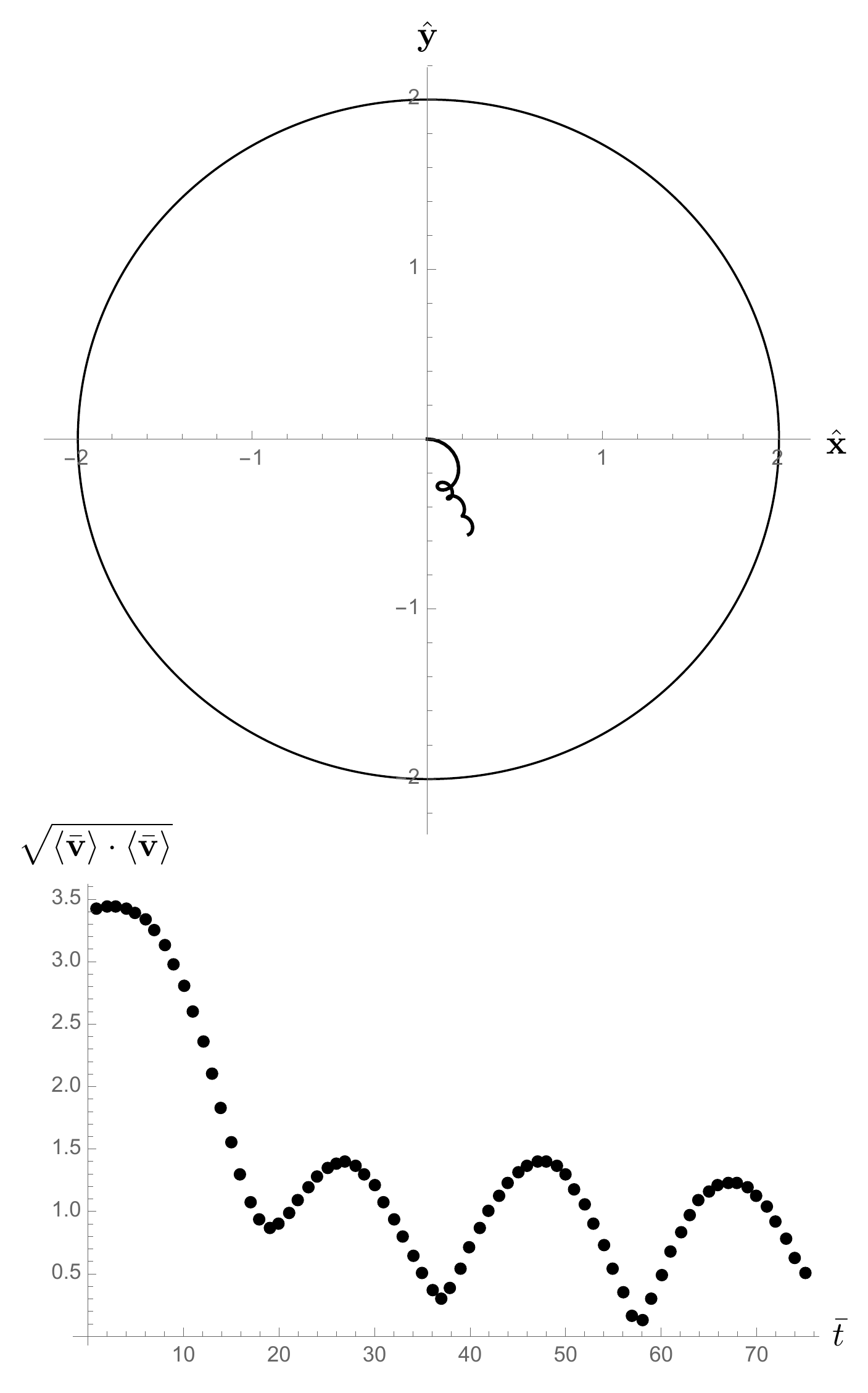} 
   \caption{An example of a ``trapped" trajectory (for the amount of time available, given boundary effects) -- the position $\langle \vecbf x \rangle$ is shown above, with the speed below.}
   \label{fig:bounded}
\end{figure}

To exhibit ``bound" behavior, we raise the height of the magnetic ``barrier", taking $\alpha = 40$ and leaving everything else the same.  The resulting trajectory is shown in the top panel of~\reffig{fig:bounded} (here we take $75$ steps) -- this time, the ``speed" of the particle is {\it not} constant (shown in the lower panel of~\reffig{fig:bounded}), yet the energy remained constant to within $10^{-12}$ (meaning the difference of the maximum value and minimum value of energy over the time-scales shown in the position expectation value plot). This is fundamentally different behavior than the classical case and comes from the fact that the notion of ``speed" in quantum mechanics has two different interpretations -- there is the magnitude of the expectation value of velocity, $\sqrt{\langle \vecbf v \rangle \cdot \langle \vecbf v \rangle}$ which is not constant, and alternatively $\sqrt{\langle \vecbf v \cdot \vecbf v \rangle}$ which is constant.  In classical mechanics, there is no distinction to be made.

\begin{figure}[htbp] 
   \centering
   \includegraphics[width=4in]{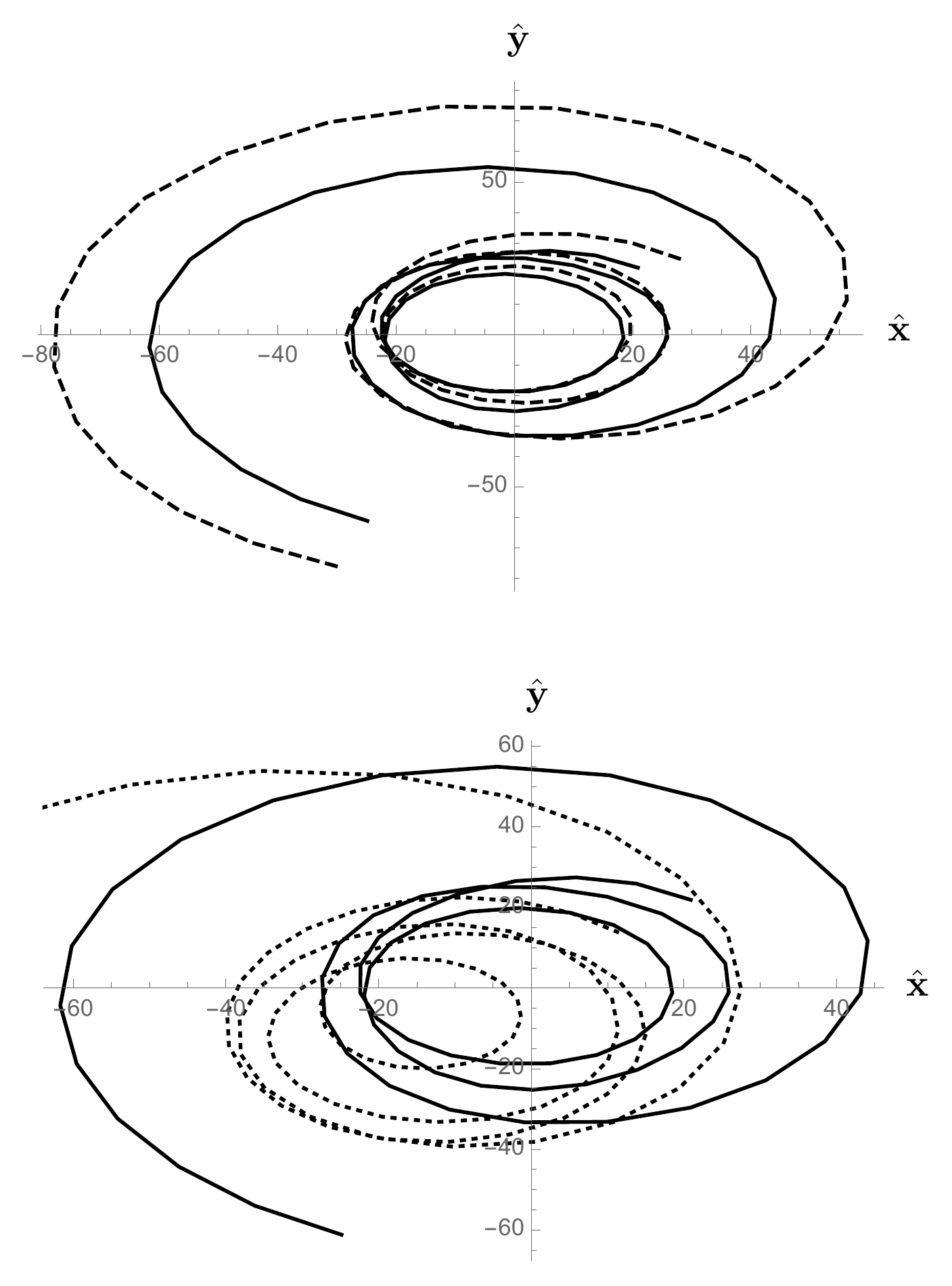} 
   \caption{The forces for the bound case (again with curves connecting the tips of the force vectors themselves) -- in the top figure, the solid line is calculated from the left-hand side of~\refeq{expp}, the dashed line is computed using the right-hand side of~\refeq{expp}.  In the bottom figure, the solid line is again the left-hand side of~\refeq{expp} the dotted line is $q \, \langle \vecbf v \rangle \times \vecbf B$, the right-hand side of~\refeq{wrongexpp}.  The two cases have been separated here for clarity.}
   \label{fig:trapforce}
\end{figure}

We can once again compare the ``forces" defined by~\refeq{expp} and~\refeq{wrongexpp} in the trapped case -- those are shown in~\reffig{fig:trapforce}, and again we see that the left and right-hand sides of~\refeq{expp} are better matched than the left-hand side of~\refeq{expp} and the fictitious $q \, \langle \vecbf v \rangle \times \vecbf B$ (the right-hand side of~\refeq{wrongexpp}).  These force expectation values introduce additional error, above and beyond the discretization error in the Crank-Nicolson method itself, because of the approximations to both derivatives and integrals needed to evaluate them, so we don't expect perfect matches.

While there are numerical errors associated both with the Crank-Nicolson method and the calculation of expectation values, there is an implicit physical difference between the quantum mechanical problem and the classical one.   Our numerical method required that the wave function vanish at the edges of our square domain, we put an infinite square well around the domain to keep the particle localized.  There is no such constraining force in the classical problem -- nor would the constraining force play much of a role there -- if we confined the classical trajectory to live in a box of side length $5 \, R$ (where $R$ is the radius of the field region), and we considered trapped motion, the boundary would never be probed.  The quantum mechanical effect of the boundary is very different -- there is non-zero probability of finding the particle outside the magnetic field region, even for cases in which the expectation value of position remains inside the field region, and that ``external" portion of the wave function reflects off of the boundary.  Because our expectation values are integrated over the entire domain, those boundary effects get transmitted to the dynamics of the expectation value.  We have attempted to minimize this contribution to our problem by placing the boundaries far away, and keeping the initial Gaussian localized within the field region -- but the boundaries do put a bound on how long we expect to be able to compare the classical and quantum trajectories.

 \section{Conclusion}
 
 The motion of particles in the presence of magnetic fields is complicated -- few closed-form solutions exist, and while we can say quite a bit about the behavior of particles moving in magnetic fields, the trajectories themselves require numerical solution, even classically.  The situation is worse quantum mechanically -- even constant magnetic fields prove difficult to handle -- solving Schr\"odinger's equation for such fields, starting from a reasonable initial wave function (like Gaussian) is not possible analytically.  In this paper, we use numerical methods to study the motion of particles in magnetic fields, both classical trajectories (solved using Runge-Kutta methods) and quantum ones using a modification of Crank-Nicolson.  We started by looking at the classical problem of particle motion, first showing that for radially symmetric flux-free fields, particles will escape the circular field region provided their initial speed is larger than the ``escape" speed set by the magnetic vector potential.  We generated some trajectories for both ``trapped" and ``escape" behavior numerically to verify that the escape speed matches its theoretical prediction.  That prediction relied on a completely gauge-fixed magnetic vector potential in Coulomb gauge -- it would be interesting to explore the effect of other gauge choices.

On the quantum mechanical side, we extended Crank-Nicolson to handle magnetic fields while retaining the norm-preservation of the method.  Using a linear, flux-free magnetic field, we
verified that the behavior of the position expectation value matches the classical trajectories in the following ways:  1.\  Particles exit perpendicular to the boundary of the field region, and 2.\  Trajectories can remain inside the field region or escape, depending on the relation of the initial momentum to the field strength~\cite{NOTE}.  We also verified that the expectation value of energy remains constant, agreeing with the classical result, and yet classically, energy conservation {\it means} that the speed of the particle is constant (since the only energy is kinetic) -- for the quantum mechanical particle, however, the speed $\sqrt{\langle\vecbf v \rangle \cdot \langle \vecbf v \rangle}$ is not constant, even though the energy is (so that $\sqrt{\langle \vecbf v \cdot \vecbf v \rangle}$ is constant).

In the case of a uniform magnetic field, our classical intuition can be used to predict the behavior of quantum mechanical expectation values, basically because the dynamical variable $\langle \vecbf v \rangle$ appears in~\refeq{wrongexpp} just as $\vecbf v$ appears in the Lorentz force law (and indeed, we recover circular motion with predictable radius and constant speed using our initial Gaussian and a constant magnetic field for $\Psi$ solved using our numerical Crank-Nicolson method).  For the more complicated flux-free magnetic field considered here, 
our classical intuition does not help us, because the effective force on the right-hand side of~\refeq{expp} involves $\vecbf p$ and $\vecbf B$ inside the expectation value -- roughly speaking, we are looking at an effective force of the form $q \langle \vecbf v \times \vecbf B \rangle$ rather than $q \, \langle \vecbf v \rangle \times \vecbf B$, {\it different} effective forces, leading to demonstrably different dynamics.   In the context of Ehrenfest's theorem (see, for example~\cite{GriffithsQM}, the informal statement is that ``quantum mechanical expectation values obey classical laws"), while~\refeq{expp} does give us a classical ``law" like Newton's second law, the force on the right is unfamiliar, and not directly comparable to the Lorentz force law.  It would be interesting to try to generate a classical analogue to the quantum mechanical effective force in~\refeq{expp} so that a direct comparison of the classical (under the influence of a modified ``effective" force) and quantum trajectories was possible~\cite{NOTENOTE}.

\acknowledgements
The authors thank David Griffiths for useful commentary and physical insight.


\begin{thebibliography}{18}
\bibitem{NOTEONE} This is an example of Ehrenfest's theorem. The proof of this type of identity comes directly from calculating the expectation value of velocity and taking the time-derivative of it, then employing Schr\"odinger's equation to remove the time-derivatives of $\psi$ that naturally appear.
\bibitem{NOTE} although unlike the classical motion, it is difficult to determine, given the magnitude of $\vecbf B$, exactly what initial momentum will lead to bound motion or escape.  It would be interesting to estimate the ``escape speed" for the quantum mechanical problem, although this is much more difficult than its classical analogue because of the different behaviors of $\sqrt{\langle \vecbf v \cdot \vecbf v \rangle}$, $\sqrt{\langle \vecbf v \rangle \cdot \langle \vecbf v \rangle }$ and $\langle E \rangle$ itself, which is not just the kinetic piece for a Gaussian initial wave function ($\langle E \rangle$ also depends on $a$ in~\refeq{initialGaussian}).
\bibitem{NOTENOTE}  There should be a transition from quantum mechanical behavior to classical in some regime (of angular momentum, say), allowing for a convergence of our two effective ``forces", and the details of that convergence would be interesting to probe.
\end{thebibliography}
\end{document}